\documentclass[12pt]{article}
\usepackage{amssymb,amsmath,amsthm,latexsym}

\newtheorem{thm}{Theorem}[section]

\newtheorem{lem}[thm]{Lemma}
\newtheorem{cor}[thm]{Corollary}
\newtheorem{exam}{Example}
\newtheorem{defi}[thm]{Definition}
\newtheorem{rem}[thm]{Remark}
\newcommand{\pf}{{\bf Proof. \ }}

\begin{document}

\title{Constructions of quantum MDS codes}
\author{Hualu Liu\footnote{Corresponding author.}\\
  School of Science, \\
 Hubei University of Technology \\
 Wuhan, Hubei 430068, China, \\
{Email: \tt hwlulu@aliyun.com} \\
 Xiusheng Liu\\
 School of Mathematics and Physics, \\
 Hubei Polytechnic University  \\
 Huangshi, Hubei 435003, China, \\
{Email: \tt lxs6682@163.com} \\}
\maketitle


\begin{abstract}
Let $\mathbb{F}_q$ be a finite field  with $q=p^{e}$ elements, where $p$ is a prime number and $e \geq 1$ is an integer.  In this paper, by means of generalized Reed-Solomon (GRS) codes, we construct two new classes of quantum maximum-distance-separable ( quantum MDS) codes with parameters
 $$[[q + 1, 2k-q-1, q-k+2]]_q$$
for $\lceil\frac{q+2}{2}\rceil \leq k\leq q+1$, and
$$[[n,2k-n,n-k+1]]_q$$
for $n\leq q $ and $ \lceil\frac{n}{2}\rceil \leq k\leq n$.  Our constructions improve and generalize some results of available in the literature. Moreover, we give an affirmative answer to the open problem proposed by Fang et al. in \cite{Fang1}.
\end{abstract}

\bf Key Words\rm :  $s$-Galois dual codes; quantum MDS codes; GRS codes

\section{Introduction}

Quantum error correction codes play an important role in the fields of fighting against decoherence and quantum noise. It is well-known that a  class of quantum error correction codes is obtained from the Calderbank-Shor-Steane (CSS) construction \cite{Calderbank1}.  According to \cite{Calderbank2}, a CSS quantum code is succinctly represented as a pair of linear codes $(C_1 ,C_2)$ over finite fields $\mathbb{F}_{p^e}$ with $C_2\subset C_1$, where $p$ is a prime and $e$ is a positive integer. Note that a CSS quantum code is a Hilbert space associated with a code pair $(C_1 ,C_2)$ in the manner described in \cite{Steane1} with $ C_2^{\perp}\subset C_1$. In this paper, any code pair written in the form  $(C_1 ,C_2)$  is supposed to satisfy the constraint $C_2^{\perp_s}\subset C_1$, where $C^{\perp_s}$ denotes the $s$-Galois dual of $C$ with $0\leq s<e$ (see \cite{Fan,Liu1}).

There are only few works dealing with CSS construction based on two distinct classical codes \cite{Aly,Aly1,Ha,La,Ma}. In \cite{Aly,Aly1}, existence conditions on quantum BCH codes are established, whereas in \cite{La} quantum codes derived from  two distinct classical BCH codes are presented. In \cite{Ha} concatenated quantum codes have been constructed. All of  CSS construction are only to BCH codes or constacyclic codes. Recently, we in \cite{Liu2} proposed two new  criteria  for  the  $C_2^{\perp_s}\subset C_1$ based on $\mathrm{rk}(G_1(G_2^{(p^{e-s})})^{T})$ or $\mathrm{rk}\begin{pmatrix}G_1\\H_2^{(p^{e-s})}\end{pmatrix}$ with $0\leq s<e$, where $G_1$ and $G_2$ are  generator matrices for $C_1$ and $C_2$, respectively, and $H_2$ is parity-check matrix for the linear code $C_2$. Then we gave a class of new  quantum codes and some new  quantum MDS codes. The aim of this paper is to continue  constructing new quantum MDS codes by the methods of ours which used in \cite{Liu2}.

This paper organized as follows. In Section 2, we recall the basic definitions and properties of linear codes, $s$-Galois dual codes and quantum  codes. In Section 3, using the methods of \cite{Liu2}, we obtain two new classes of MDS quantum  codes by means of Generalized Reed-Solomon (GRS) codes, one of which generalizes a result of \cite{Liu2}.  Finally, a brief summary of this work is described in Section 4.

\section{Preliminaries}
In this section, we recall some basic concepts and results about linear codes, $s$-Galois dual codes and quantum codes, necessary for the development of this work. For more details, please refer to \cite{Fan,Huffman, Liu1,Aly}.

Throughout this paper, let $\mathbb{F} _{q}$ be a finite field with $q=p^{e}$ elements, where $p$ is a prime number and $e \geq 1$ is an integer. For a positive integer $n$, let $\mathbb{F}_{q}^{n}$
denote the vector space of all $n$-tuples over $\mathbb{F} _{q}$. A linear $[n,k]_{q}$ code $C$ over $\mathbb{F} _{q}$ is  a $k$-dimensional subspace of $\mathbb{F}_{q}^{n}$. The Hamming weight $w_H(\mathbf{c})$ of a codeword $\mathbf{c} \in C$ is the number of nonzero components of $\mathbf{c}$. The Hamming distance of two codewords $\mathbf{c}_1,\mathbf{c}_2 \in C$  is $d_H(\mathbf{c}_1,\mathbf{c}_2) = w_H(\mathbf{c}_2-\mathbf{c}_1)$. The minimum Hamming distance $d$ of $C$ is the minimum Hamming distance between any two distinct codewords of $C$. An $[n,k,d]_{q}$ code is an $[n,k]_{q}$ code with the minimum Hamming distance $d$.

\subsection{$s$-Galois dual codes}
In \cite{Fan}, Fan and Zhang introduce a kind of forms on $\mathbb{F}_{q}^n$
as follows (\cite[Definition 4.1]{Fan}).

For each integer $s$ with $0\leq s < e $ , define:
$$
[{\bf x},{\bf y}]_{s}=x_1y_1^{p^{s}}+\cdots+x_ny_n^{p^{s}},
\qquad\forall~ {\bf x},{\bf y}\in\mathbb{F}_{q}^n.
$$
We call $[{\bf x},{\bf y}]_{s}$ the $s$-Galois form on $\mathbb{F}_{q}^n$.
It is just the usual Euclidean inner product if $s=0$. And, it is  the Hermitian inner product if $e$ is even and $s=\frac{e}{2}$.
For any code $C$ of $\mathbb{F}_{q}^n$, the following code
$$
C^{\bot_{s}}=\big\{{\bf x}\in\mathbb{F}_{q} ^n\,\big|\,
 [{\bf c},{\bf x}]_{s}=0,\forall~{\bf c}\in C\big\}
$$
is called the $s$-Galois dual code of $C$.
Note that $C^{\bot_{s}}$ is linear whenever $C$ is linear or not.
Then $C^{\bot_{0}}$ (simply, $C^{\perp}$) is just the Euclidean dual code of $C$,
and $C^{\bot_{\frac{e}{2}}}$ (simply, $C^{\perp_{H}}$)
is just the Hermitian dual code of $C$.
If $C \subset C^{\bot_{s}}$, then $C$ is said to be $s$-Galois self-orthogonal.
Moreover, $C$ is said to be $s$-Galois self-dual if $C= C^{\bot_{s}}$.

From the fact that $s$-Galois form is non-degenerate (\cite[Remark 4.2]{Fan}),
it follows immediately that
$\dim_{\mathbb{F}_{q}}C+\dim_{\mathbb{F}_{q}}C^{\perp_{s}}=n$.

A  linear code $C$ is called $s$-Galois dual-containing if $C^{\perp_{s}}\subset C$.

For a $t\times t$ matrix $A =(a_{ij})_{t\times t}$ over $\mathbb{F}_{q}$,
we denote $A^{(p^{e-s})} = (a_{ij}^{p^{e-s}})_{t\times t}$.
In particular, for a vector $\mathbf{a}=(a_1,a_2,\ldots,a_n)\in\mathbb{F}_{q} ^n$,
we have
$$\mathbf{a}^{p^{e-s}}=(a_1^{p^{e-s}},a_2^{p^{e-s}},\ldots,a_n^{p^{e-s}}).$$
And for a linear code $C$ of $\mathbb{F} _{q}^n$,
we define $C^{p^{e-s}}$ to be the set
$\{\mathbf{a}^{p^{e-s}}\mid~\mathbf{a}\in C\}$ which is also a linear code.
It is easy to see that the $s$-Galois dual $C^{\perp_{s}}$ is equal to the
Euclidean dual $(C^{p^{e-s}})^{\perp}$ of the linear code $C^{p^{e-s}}$.

\subsection{Quantum codes}
Let $V_n$ be the Hilbert space $V_n=\mathbb{C}^{q^n}=\mathbb{C}^{q}\otimes\cdots\otimes\mathbb{C}^{q}$. Let $|x\rangle$ be the vectors of an orthonormal basis of $\mathbb{C}^{q^n}$, where the labels $x$ are elements of $\mathbb{F}_{q}$. Then $V_n$ has the following orthonormal basis $\{|\mathbf{c}\rangle=|c_1c_2\cdots c_n\rangle=|c_1\rangle \otimes|c_2\rangle \otimes\cdots\otimes |c_n\rangle:~\mathbf{c}=(c_1,c_2,\ldots,c_n) \in \mathbb{F}_{q}^{n}\}$.

Consider $a,b\in \mathbb{F}_{q}$, the unitary linear operators $X(a)$ and $Z(b)$ in $\mathbb{C}^{q}$ are defined by $X(a)|x\rangle=|x+a\rangle$ and $Z(b)|x\rangle=\omega^{{\rm tr}(bx)}|x\rangle$, respectively, where $\omega=exp(2\pi i/p)$ is a primitive $p$-th root of unity and ${\rm tr}$ is the trace map from  $\mathbb{F}_{q}$ to  $\mathbb{F}_{p}$.

Let $\mathbf{a}=(a_1,\ldots,a_n)\in \mathbb{F}_{q}^n$, we write $X(\mathbf{a})=X(a_1)\otimes\cdots\otimes X(a_n)$ and $Z(\mathbf{a})=Z(a_1)\otimes\cdots\otimes Z(a_n)$ for the tensor products of $n$ error operators. The set $E_n=\{X(\mathbf{a})Z(\mathbf{b}): \mathbf{a},\mathbf{b}\in \mathbb{F}_{q}^n\}$ is an error basis on the complex vector space $\mathbb{C}^{q^n}$ and then we set $G_n=\{\omega^cX(\mathbf{a})Z(\mathbf{b}): \mathbf{a},\mathbf{b}\in \mathbb{F}_{q}^n,c\in\mathbb{F}_{p} \}$ is the error group associated with $E_n$.

\begin{defi}\label{de:2.1}
 A $q$-qry quantum code of length $n$ is a subspace $Q$ of $V_n$ with dimension $K>1$.  A quantum code $Q$ of dimension $K>2$ is called  quantum code  with parameters $((n,K,d))_q$ or $[[n,k,d]]_q$, where $k=log_qK$ if $Q$ detect $d-1$ quantum digits of errors for $d\geq1$. Namely, if for every orthogonal pair $|u\rangle,~|v\rangle$ in $Q$ with $<u|v>=0$ and every $e\in G_n$ with $W_Q(e)\leq d-1$, $|u\rangle$ and $e|v\rangle$ are orthogonal, i.e., $<u|e|v>=0$. Such a quantum code is called pure if $<u|e|v>=0$ for any  $|u\rangle$ and $|v\rangle$  in $Q$ and any $e\in G_n$ with $1\leq W_Q(e)\leq d-1$. A quantum code $Q$ with $K=1$ is always pure.
\end{defi}

From the classical linear codes,  we can directly obtain a family of quantum codes by using the called CSS given by the
following theorem.
\begin{thm}\label{th:2.2}(\cite{Calderbank,Calderbank1,Steane1}). (CSS Code Construction) Let $C_1$ and $C_2$ denote two classical linear codes with parameters $[ n , k_1 , d_1 ]_q$ and
$[ n , k_2 , d_2 ]_q$, respectively, such that $C_2^{\perp} \subset C_1$.Then there exists an $[[n,k_1+k_2-n, d ]]_q$ quantum code where $d= \mathrm{min}\{W_H( c )|c\in( C_1\backslash C_2^{\perp} )\cup ( C_2\backslash C_1^{\perp}\}$ that is pure to $\mathrm{min}\{d_1 ,d_2 \}$.
\end{thm}

\begin{cor}\label{co:2.3} Let $C_1$ and $C_2$ denote two classical linear codes with parameters $[ n , k_1 , d_1 ]_q$ and $[ n , k_2 , d_2 ]_q$, respectively, such that $C_2^{\perp_s} \subset C_1$. Then there exists an $[[n,k_1+k_2-n, d ]]_q$ quantum code where $d= \mathrm{min}\{W_H( c )|c\in( C_1\backslash (C_2^{p^{e-s}})^{\perp} )\cup ( C_2^{p^{e-s}}\backslash C_1^{\perp})\}\geq\mathrm{min}\{d_1,d_2\}$ that is pure to $\mathrm{min}\{d_1 ,d_2 \}$.
\end{cor}
\pf Obviously, $C_2^{p^{e-s}}$ is also an $[n,k_2,d_2]_q$ linear code over $\mathbb{F}_q$. Since $C_2^{\perp_s} \subset C_1$ is equal to $C_1^{\perp}\subset C_2^{p^{e-s}}$, there exists an $[[n,k_1+k_2-n, d ]]_q$ quantum code where $d= \mathrm{min}\{W_H( c )|c\in( C_1\backslash (C_2^{p^{e-s}})^{\perp} )\cup ( C_2^{p^{e-s}}\backslash C_1^{\perp}\}$ by Theorem \ref{th:2.2}.
\qed

\begin{lem}\label{le:B}
Let $C_i$ be an $[n,k_i]_{q}$ linear code over $\mathbb{F}_{q}$ with generator matrix $G_i$ for $i=1,2$.  Then $~\mathrm{rk}(G_1(G_2^{(p^{e-s})})^{T})=k_1- \mathrm{dim}_{F_q}(C_1\cap C_2^{\perp_s})$.
\end{lem}

The following two lemmas  characterizes the $s$-Galois dual-containing of two linear codes, the proofs of which can be found in \cite{Liu2}.
\begin{lem}\label{le:2.1}
Let $C_i$ be an $[n,k_i]_{q^{2}}$ linear code over $\mathbb{F}_{q}$ with generator matrix $G_i$ for $i=1,2$. Then $C_2^{\perp_{H}}\subset C_1$ if and only if   $~\mathrm{rk}(G_1(G_2^{(p^{e-s})})^{T})\leq k_1+k_2-n~$ and $~k_1+k_2\geq n$.
\end{lem}

\begin{lem}\label{le:2.4} Let $C_i$ be an $[n,k_i]_{q}$ linear code over $\mathbb{F}_{q}$ with generator matrix $G_i$ and parity-check $H_i$ for $i=1,2$.  Then
$C_2^{\perp_{s}}\subset C_1$ if and only if  $\mathrm{rk}\begin{pmatrix}G_1\\H_2^{(p^{e-s})}\end{pmatrix}\leq k_1$ and $k_1+k_2\geq n$.
\end{lem}

Combining Lemma \ref{le:2.1} or Lemma \ref{le:2.4} and Corollary \ref{co:2.3}, we have the following  new methods to construct quantum codes.

\begin{thm}\label{th:2.6}
Let $C_i$ be an $[n,k_i]_{q}$ linear code over $\mathbb{F}_{q}$ with generator matrix $G_i$ or parity-check matrix $H_i$ for $i=1,2$. If   $~\mathrm{rk}\begin{pmatrix}G_1\\H_2^{(p^{e-s})}\end{pmatrix}\leq k_1~$ or $~\mathrm{rk}(G_1(G_2^{(p^{e-s})})^{T})=k_1-\mathrm{dim}_{F_q}(C_1\cap C_2^{\perp_s})\leq k_1+k_2-n~$ and $~k_1+k_2\geq n$, then there exist a quantum code with parameters $[[n,k_1+k_2-n, d]]_{q}$, where $d\geq\mathrm{min}\{d_1,d_2\}$.
\end{thm}

To see that a quantum code $Q$ is good in terms of its parameters, we have to introduce the quantum Singleton bound (see \cite{Ashikhmin}).
\begin{thm}\label{th:2.7}
Let $Q$ be a quantum code with parameters $[[n,k,d]]_{q}$. Then $2d\leq n-k+2$.
\end{thm}

If  a quantum code  $Q$ with parameters $[[n,k,d]]_{q}$ attains the quantum Singleton bound $2d= n-k+2$, then it is called a quantum  maximum-distance-separable (MDS) code.


\section{ New quantum  MDS codes construction}
In this section, we construct  two new classes of quantum MDS codes.
\subsection{Codes construction I}
The hull of $C$ is the code $C\cap C^{\perp}$, denoted by $\mathrm{Hull}(C)$, in the terminology that was introduced
in \cite{Jr}.

The following two lemmas is due to Luo, Cao and Chen \cite{Luo}.

\begin{lem}\label{le:3.1} Let $q > 3$ be an odd prime power. Then

$(1)$ there exists an $[q + 1, k, q-k+2]_q$ MDS code with $l$-dimensional hull for any $1\leq l\leq k - 1 $.

$(2)$ there exists an $[q + 1,\frac{q + 1}{2}, \frac{q + 1}{2}+1]_q$ MDS code
with $l$-dimensional hull for any $1 \leq l \leq \frac{q + 1}{2}$.
\end{lem}

\begin{lem} \label{le:3.2}Let $q =2^m$, where $m > 1$ is an integer. Then there exists an $[q + 1, k, q-k+2]_q$ MDS code
with $l$-dimensional hull for any $1\leq l\leq k - 1 $.
\end{lem}

With the above lemmas and Theorem \ref{th:2.6}, the construction of MDS quantum codes turns into that of MDS linear codes
with the determined dimensional hull. Then we have the following results.

\begin{thm} \label{th:3.4}Let $q > 3$ be an odd prime power.

$(1)$~If  $\lceil\frac{q+2}{2}\rceil \leq k\leq q+1$, then there exists an $[[q + 1, 2k-q-1, q-k+2]]_q$ quantum MDS code.

$(2)$ There exists an $[[q + 1,0, \frac{q + 1}{2}+1]]_q$ quantum MDS code.
\end{thm}
\pf $(1)$ In Lemma \ref{le:3.1}$(1)$, let $q+1-k\leq l \leq k-1$. Then $k\geq \lceil\frac{q+2}{2}\rceil$. By Lemma \ref{le:B}, we have
$$\mathrm{rank}(GG^T) = k-l\leq 2k-q-1.$$
According  to Theorems  \ref{th:2.6} and  \ref{th:2.7},  there exists an $[[q + 1, 2k-q-1, q-k+2]]_q$ quantum MDS code.

$(2)$ In Lemma \ref{le:3.1}$(2)$, let $ l=k=\frac{q+1}{2}$.  By Lemma \ref{le:B}, we have
$$\mathrm{rank}(GG^T) = k-l=0\leq2k-q-1.$$
According  to Theorems  \ref{th:2.6} and  \ref{th:2.7},  there exists an $[[q + 1, 0, \frac{q + 1}{2}+1]]_q$ quantum MDS code.
\qed

The proof of the following theorem is similar to the above theorem  and is omitted here.

\begin{thm}  \label{th:3.5} Let $q =2^m$, where $m > 1$ is an integer. If $ \frac{q+2}{2}\leq k\leq q+1$, then there exists an $[[q + 1, 2k-q-1, q-k+2]]_q$ quantum MDS code.
\end{thm}

\begin{rem} Fang et al.\cite{Fang1} gave an open problem: if $l=2^s>2$, does there exist a quantum MDS code of length $l^2+1$ and minimum distance $l$ ? In Theorem \ref {th:3.5}, let $m=2s$ such that $q=l^2$. Taking $k=\frac{l^2+2}{2}+\frac{(l-1)^2+1}{2}$,  then there exists an $[[l^2 + 1, l^2-2l+3, l]]_q$ quantum MDS code, which is implies that the open problem is right.
\end{rem}

\begin{exam} In Theorems \ref{th:3.4} and \ref{th:3.5}, taking some special values of $q$, we obtain some
quantum MDS codes in Table $1$. In fact, Theorems \ref{th:3.4} and \ref{th:3.5}  include the Theorem 4.2 given in \cite{Jin}.
\end{exam}
\begin{table}
\caption{New quantum MDS codes}
\begin{center}\begin{tabular}{c|c|c}
\hline
$q$&New quantum MDS codes & quantum MDS codes from Theorem 4.2 in \cite{Jin}\\
\hline
$9$&$[[10,2,5]]_{9},[[10,4,4]]_{9},[[10,6,3]]_{9}$,&$[[10,8,2]]_{9},[[10,6,3]]_{9},[[10,4,4]]_{9}$,\\&$[[10,8,2]]_{9},[[10,10,1]]_{25},[[10,0,6]]_{9}$\\
\hline
$16$&$[[17,1,9]]_{16},[[17,3,8]]_{16},[[17,5,7]]_{16}$,&$[[17,17,1]]_{16},[[17,15,2]]_{16},[[17,13,3]]_{16}$,\\&$[[17,7,6]]_{16},[[17,9,5]]_{16},[[17,11,4]]_{16}$,&$[[17,11,4]]_{16},[[17,9,5]]_{16} $,\\&$[[17,13,3]]_{16},[[17,15,2]]_{16},[[17,17,1]]_{16}$\\
\hline
25&$[[26,2,13]]_{25},[[26,4,12]]_{25},[[26,6,11]]_{25}$,&$[[26,16,6]]_{25},[[26,18,5]]_{25},[[26,20,4]]_{25}$,\\&$[[26,8,10]]_{25},[[26,10,9]]_{25},[[26,12,8]]_{25}$ &$[[26,22,3]]_{25},[[26,24,2]]_{25},[[26,26,1]]_{25}$
\\&$[[26,14,7]]_{25},[[26,16,6]]_{25},[[26,18,5]]_{25}$,\\& $[[26,20,4]]_{25},[[26,22,3]]_{25},[[26,24,2]]_{25}$,\\&$[[26,26,1]]_{25},[[26,0,14]]_{25}$\\
\hline
\end{tabular}\end{center}
\end{table}

\subsection{Codes construction II}

Firstly, we review some basic notations and results about generalized Reed-Solomon codes. For the details, the reader is referred to \cite{Jin2}. Let  $n$ be a positive integer with $1 <n \leq q$, $\alpha_1,\ldots,\alpha_n$ be $n$ distinct elements of $\mathbb{F}_{q}$, and let  $v_1,\ldots,v_n$ be $n$ nonzero elements of $\mathbb{F}_{q}$. For $k$ between $1$ and $n$, the generalized Reed-Solomon code $GRS_k(\mathbf{a},\mathbf{v})$ is defined by
$$GRS_k(\mathbf{a},\mathbf{v})=\{(v_1f(\alpha_1),\ldots,v_nf(\alpha_n))|~f(x)\in\mathbb{F}_{q}[x],deg(f(x))\leq k-1\},$$
where $\mathbf{a}$, $\mathbf{v}$ denote the vectors $(\alpha_1,\ldots,\alpha_n)$, $(v_1,\ldots,v_n)$, respectively.
Clearly, $GRS_k(\mathbf {a},\mathbf{v})$ has a generator matrix
$$G=\begin{pmatrix}v_1&v_2&\cdots&v_n\\v_1\alpha_1&v_2\alpha_2&\cdots&v_n\alpha_n\\v_1\alpha_1^2&v_2\alpha_2^2&\cdots&v_n\alpha_n^2\\\vdots&\vdots&\cdots&\vdots\\v_1\alpha_1^{k-1}&v_2\alpha_2^{k-1}&\cdots&v_n\alpha_n^{k-1}\end{pmatrix}.$$
It is well known that the code $GRS_k(\mathbf{a},\mathbf{v})$ is an $[n, k, n- k + 1]_q$ MDS code.

The following  result can be found in \cite{Jin2}.
\begin{lem}\label{le:3.5} Let $\mathbf{1}$  be all-one word of length $n$.   The dual code of $GRS_k(\mathbf{a},\mathbf{1})$ is $GRS_{n-k}(\mathbf{a},\mathbf{u})$,
where $\mathbf{u} = (u_1, u_2, \ldots , u_n)$ with $u_i = \prod_{1\leq j\leq n,j\neq i}(\alpha_i-\alpha_j)^{-1}$.
\end{lem}

\begin{thm}\label{th:3.7} Let $n\leq q$, $1\leq k_1\leq n$ and $1\leq k_2\leq n$. If $k_1+k_2\geq n$, then

$(1)$ there exists a quantum code with parameters $[[n,k_1+k_2-n,d]]_q$, where $d\geq \mathrm{min} \{n-k_1+1,n-k_2+1\}$;

$(2)$ when $k_1=k_2=k$ and $k\geq \lceil\frac{n}{2}\rceil$, there exists a quantum MDS code with parameters $[[n,2k-n,n-k+1]]_q$.
\end{thm}
\pf $(1)$ Let $\mathbf{u} = (u_1, u_2, \ldots , u_n)$, where $u_i = \prod_{1\leq j\leq n,j\neq i}(\alpha_i-\alpha_j)^{-1}$  for $1\leq i\leq n$.
It is well known that $GRS_{k_1}(\mathbf{a},\mathbf{u})$ has a generator matrix
$$G_{1}=\begin{pmatrix}u_1&u_2&\cdots&u_n\\u_1\alpha_1&u_2\alpha_2&\cdots&u_n\alpha_n\\u_1\alpha_1^2&u_2\alpha_2^2&\cdots&u_n\alpha_n^2\\
\vdots&\vdots&\cdots&\vdots\\u_1\alpha_1^{k_1-1}&u_2\alpha_2^{k_1-1}&\cdots&u_n\alpha_n^{k_1-1}\end{pmatrix} .$$

By Lemma \ref{le:3.5}, $GRS_{k_2}(\mathbf{a},\mathbf{1})$  has a parity check matrix
$$H_{2}=\begin{pmatrix}u_1&u_2&\cdots&u_n\\u_1\alpha_1&u_2\alpha_2&\cdots&u_n\alpha_n\\u_1\alpha_1^2&u_2\alpha_2^2&\cdots&u_n\alpha_n^2\\
\vdots&\vdots&\cdots&\vdots\\u_1\alpha_1^{n-k_2-1}&u_2\alpha_2^{n-k_2-1}&\cdots&u_n\alpha_n^{n-k_2-1}\end{pmatrix}.$$

If $k_1+k_2\geq n$, i.e., $k_1-1\geq n-k_2-1$, then  $\mathrm{rk}\begin{pmatrix}G_1\\H_2^{(p^{e-0})}\end{pmatrix}=k_1\leq k_1$. Thus, by Theorem \ref{th:2.6}, there exists a quantum code with parameters $[[n,k_1+k_2-n,d]]_q$, where $d\geq \mathrm{min} \{n-k_1+1,n-k_2+1\}$.

$(2)$ When $k_1=k_2=k$, then by $k_1+k_2\geq n$, we have $k\geq \lceil\frac{n}{2}\rceil$. Note that $d\geq \mathrm{min} \{n-k_1+1,n-k_2+1\}=n-k+1$.

On the other hand, by Theorem \ref{th:2.7}, $2d\leq  n-(2k-n)+2=2(n-k+1)$, i.e., $d\leq n-k+1$.

It follows that $d=n-k+1$, which implies that there exists a quantum MDS code with parameters $[[n,2k-n,n-k+1]]_q$.
\qed
\begin{rem} Theorem \ref{th:3.7} generalizes a result of \cite{Liu2}. In the above Theorem \ref{th:3.7}, when $n |q-1$ and $n-k=1+j$ we can obtain Theorem 4.11 in  \cite{Liu2}.
\end{rem}

\begin{exam} In Theorem \ref{th:3.7}, let $q=l^2$ where $l$ is a prime power.  In Table 2, we compare our quantum MDS codes with previously known quantum MDS codes available in \cite{Fang},\cite{Fang1},\cite{Jin1} and  \cite{Kai}.  As can be seen, when the code length $n$ is fixed, the new quantum MDS codes have much larger minimum distance.
\end{exam}
\begin{table}
\caption{ A comparison of new quantum MDS codes}
\begin{center}\begin{tabular}{c|c|c|c}
\hline
$q$ & Length $n$& Distance~$d$&Distance~$d'$ in Refs.   \\
\hline
$l^2$&$n=l^2$&$2\leq d \leq l^2-\lfloor\frac{l^2}{2}\rfloor+1$ & $3\leq d'\leq l+1$ ~\cite{Jin1} \\
\hline
$l^2$&$n=l^2-1$&$2\leq d \leq l^2-\lfloor\frac{l^2-1}{2}\rfloor$ & $3\leq d'\leq l-1$ ~\cite{Jin1}\\
\hline
$l^2$&$n=1+\frac{r(l^2-1)}{s}$ &$2\leq d \leq a-b+2$\\& $s\mid (l-1)$ and $1\leq r \leq s$&$a=\frac{r(l^2-1)}{s}, b=\lfloor\frac{s+r(l^2-1)}{2s}\rfloor$ & $2\leq d'\leq \frac{r(l-1)}{s}+1$~\cite{Fang}\\
\hline
$l^2$&$n=tl,1\leq t \leq l$&$2\leq d \leq tl-\lfloor\frac{tl}{22}\rfloor+1$ & $2\leq d'\leq \lfloor\frac{tl+l-1}{l+1}\rfloor+1$~\cite{Fang1}\\
\hline
$l^2$&$n=\lambda(l+1)$, $\lambda$ is an odd \\&divisor of $l-1$&$2\leq d \leq \frac{\lambda(l+1)}{2}+1$ & $2\leq d'\leq \frac{l+1}{2}+\lambda$~\cite{Kai}\\
\hline
\end{tabular}\end{center}
\end{table}

\section{Conclusion}
In this paper, by using two methods and generalized Reed-Solomon codes, we have constructed  two new families of quantum MDS codes with flexible parameters. We see that a result of \cite {Liu2} is special case of ours and the parameters of some previous results are also improved.  Moreover, we give an affirmative answer to the open problem raised by Fang et al. in \cite{Fang1}.

\textbf{Acknowledgements}
This work was supported by Scientific Research
Foundation of Hubei Provincial Education Department of China(Grant No. Q20174503) and the National Science Foundation of Hubei
Polytechnic University of China (Grant No.12xjz14A and 17xjz03A).

\end{document}